\documentclass[letterpaper,10 pt,conference,dvipsnames]{ieeeconf}
\IEEEoverridecommandlockouts

\let\labelindent\relax

\usepackage[absolute,overlay]{textpos}
\usepackage{times}
\usepackage{enumitem}

\usepackage[usenames,dvipsnames]{xcolor}

\usepackage{amsmath,amssymb,mathtools,nccmath}

\usepackage{amsthm}
\usepackage{bbold,dsfont,bbm}
\allowdisplaybreaks

\theoremstyle{definition}

\usepackage{graphicx}
\usepackage{tikz,tikz-cd,pgfplots}
\pgfplotsset{compat=1.17}
\usepackage{subfig}

\usepackage[hidelinks]{hyperref}
\usepackage[capitalise,noabbrev]{cleveref}
\usepackage{cite}
\usepackage{todonotes}

\usepackage{multirow}
\usepackage{booktabs}
\usepackage[acronym]{glossaries}

\usepackage{comment}

\usepackage[per-mode=symbol]{siunitx}
\DeclareSIUnit{\eur}{\euro}
\DeclareSIUnit{\usd}{USD}
\DeclareSIUnit{\mph}{mph}
\DeclareSIUnit{\month}{month}
\DeclareSIUnit{\year}{year}
\DeclareSIUnit{\million}{Mil}
\DeclareSIUnit{\mile}{mile}
\DeclareSIUnit{\car}{car}
\DeclareSIUnit{\train}{train}
\DeclareSIUnit{\mmveh}{\text{$\mu$}MV}
\DeclareSIUnit{\nounit}{-}
\usepackage[font=footnotesize]{caption}
\definecolor{lightblue}{rgb}{0.60784,0.76078,0.90196}
\definecolor{darkblue}{rgb}{0.26667,0.44706,0.76863}
\definecolor{lightgreen}{rgb}{0.66275,0.81569,0.55686}
\definecolor{darkgreen}{rgb}{0.43922,0.67843,0.27843}
\definecolor{orange}{rgb}{0.92941,0.49020,0.19216}
\definecolor{yellow}{rgb}{1.00000,0.75294,0.00000}
\definecolor{grey}{rgb}{0.64706,0.64706,0.64706}
\definecolor{purple}{rgb}{0.51373,0.23529,0.04706}
\newacronym{abk:amod}{AMoD}{Autonomous Mobility-on-Demand}

\newacronym{abk:GUI}{GUI}{Graphical User Interface}
\newacronym{abk:iamod}{\mbox{I-AMoD}}{intermodal \gls{abk:amod}}
\newacronym{abk:av}{\mbox{AV}}{autonomous vehicle}
\newacronym{abk:sv}{\mbox{SV}}{standard vehicle}

\newacronym{abk:br}{BR}{Best Response}
\newacronym{abk:bpr}{BPR}{Bureau of Public Roads}
\newacronym{abk:bev}{BEV}{Battery Electric Vehicle}
\newacronym{abk:hev}{HEV}{Hybrid Electric Vehicle}
\newacronym{abk:ca}{CA}{congestion-aware}
\newacronym{abk:cara}{CARS}{congestion-aware routing scheme}
\newacronym{abk:cpo}{CPO}{complete partial order}
\newacronym{abk:cdp}{CDP}{co-design problem}
\newacronym{abk:cdpi}{CDPI}{co-design problem with implementation}
\newacronym{abk:dp}{DP}{design problem}
\newacronym{abk:dpi}{DPI}{design problem with implementation}
\newacronym{abk:dcpo}{DCPO}{directed complete partial order}
\newacronym{abk:es}{ES}{e-scooter}
\newacronym{abk:ffcs}{FFCS}{free floating car sharing systems}
\newacronym{abk:ghg}{GHG}{greenhouse gas}
\newacronym{abk:icev}{ICEV}{Internal Combustion Engine Vehicle}
\newacronym{abk:kpi}{KPIs}{Key Performance Indicators}
\newacronym{abk:lw}{LW}{Lightweight}
\newacronym{abk:mm}{{$\mu$}M}{micromobility}
\newacronym{abk:mod}{MoD}{Mobility-on-Demand}
\newacronym{abk:msp}{MSP}{Mobility Service Provider}
\newacronym{abk:mcdp}{MCDP}{Monotone Co-Design Problem}
\newacronym{abk:mcfp}{MCFP}{multi-commodity flow problem}
\newacronym[plural=NE,firstplural=Nash Equilibria (NE)]{abk:ne}{NE}{Nash Equilibrium}
\newacronym{abk:nyc}{NYC}{New York City}
\newacronym{abk:poset}{poset}{partially ordered set}
\newacronym{abk:sb}{SB}{shared bike}
\newacronym{abk:spp}{SPP}{shortest path problem}
\newacronym{abk:kdspp}{k-dSPP}{k-disjoint \gls{abk:spp}}
\newacronym{abk:su}{SU}{Sport Utility}
\newacronym{abk:SE}{SE}{Stackelberg Equilibrium}
\newacronym{abk:sdp}{SDP}{short-distance price}
\newacronym{abk:ldp}{LDP}{long-distance price}

\newcommand{\true}[1]{\texttt{T}}

\usepackage{ifxetex}
\ifxetex
\usepackage[tracking=false,kerning=false,spacing=false]{microtype}
\usepackage[protrusion=true,tracking=false,kerning=false,spacing=false]{microtype}
\else
\usepackage[tracking=false,kerning=true,spacing=true]{microtype}
\usepackage[inkscapelatex=false]{svg}

\fi
\usepackage{enumitem}
\renewcommand{\baselinestretch}{0.94}
\renewcommand{\baselinestretch}{0.99}
\theoremstyle{definition}

\newtheorem*{assumption*}{Assumption}
\newtheorem{theorem}{Theorem}

\newtheorem{definition}{Definition}

\theoremstyle{remark}
\newtheorem*{remark}{Remark}

\Crefname{figure}{Fig.}{Figures}
\makeatletter
    \if@cref@capitalise
        \crefname{subsection}{Section}{Sections}
        \crefname{subsubsection}{Section}{Sections}
        \crefname{assumption}{Assumption}{Assumptions}
        \crefname{problem}{Problem}{Problems}
    \else
        \crefname{subsection}{section}{sections}
        \crefname{subsubsection}{section}{sections}
        \crefname{assumption}{assumption}{assumptions}
        \crefname{problem}{problem}{problems}
    \fi
\makeatother

\usetikzlibrary{positioning,intersections,hobby,patterns,calc,decorations.pathmorphing,decorations.markings, shadows,shapes}
\tikzstyle{block} = [draw, rectangle, minimum height=2em, minimum width=3em,font=\bfseries,rounded corners,thick]
\tikzstyle{block1} = [draw, rectangle, minimum height=1.5em, minimum width=2.5em]
\tikzstyle{blockDyn} = [draw, rectangle, minimum height=2.5em, minimum width=3.5em, align=center, inner sep=10pt, thick, fill=white, copy shadow={draw=black,fill=black,opacity=1,shadow xshift=0.5ex,shadow yshift=-0.5ex}]
\tikzstyle{blockAlg} = [draw, rectangle, minimum height=1.5em, minimum width=2.5em, align=center, inner sep=10pt, thick]
\tikzstyle{sum} = [draw,circle]

\tikzstyle{nodePre} = [circle, draw,inner sep=1pt,node contents={$\preceq$},thick]
\tikzstyle{nodePreEmpty} = [circle, draw,inner sep=1pt,thick]
\tikzstyle{nodePos} = [circle, draw,inner sep=1pt,node contents={$\posceq$},thick]
\tikzstyle{nodeProd} = [rectangle, draw,inner sep=4pt,node contents={$\times$},rounded corners,thick]
\tikzstyle{nodeSum} = [rectangle, draw,inner sep=4pt,node contents={$\mathbf{+}$},rounded corners,thick]

\definecolor{DPgreen}{rgb}{0.0, 0.5, 0.0}
\definecolor{red}{rgb}{0.75, 0.0, 0.0}

\newif\ifmargincomments %A quick way of turning off margin comments for, say, arXiv submission
\margincommentstrue

\newif\ifextendedversion %A quick way of turning off appendix
\ifmargincomments

\else

\fi

\title{
\textbf{User-Friendly Game-Theoretic Modeling and Analysis of\\ Multi-Modal Transportation Systems}
}
\author{Margarita Zambrano, Xinling Li, Riccardo Fiorista, Gioele Zardini
\thanks{
Authors are with the Laboratory for Information and Decision Systems, and the Department of Civil and Environmental Engineering, Massachusetts Institute of Technology, USA, {\tt \{mbcz10,xinli831,fiorista,gzardini\}@mit.edu}.}
}

\begin{document}

%\begin{textblock*}{\textwidth}(15mm,18mm) % {block width} (coords) 
%\bf \textcolor{NavyBlue}{To appear in the Proceedings of the 2023 IEEE 26th International %Conference on Intelligent Transportation Systems}
%\end{textblock*}

\maketitle
%\thispagestyle{plain}
%\pagestyle{plain}
%%%%%%%%%%%%%%%%%%%%%%%%%%%%%%%%%%%%%%%%%%%%%%%%%%%%%%%%%%%%%%%%%%%%%%%%%%%%%%%%
\begin{abstract}
The evolution of existing transportation systems, mainly driven by urbanization and increased availability of mobility options, such as private, profit-maximizing ride-hailing companies, calls for tools to reason about their design and regulation.
To study this complex socio-technical problem, one needs to account for the strategic interactions of the stakeholders involved in the mobility ecosystem. 
In this paper, we present a game-theoretic framework to model multi-modal mobility systems, focusing on municipalities, service providers, and travelers.
Through a user-friendly, Graphical User Interface, one can visualize system dynamics and compute equilibria for various scenarios.
The framework enables stakeholders to assess the impact of local decisions (e.g., fleet size for services or taxes for private companies) on the full mobility system.
Furthermore, this project aims to foster STEM interest among high school students (e.g., in the context of prior activities in Switzerland, and planned activities with the MIT museum).
This initiative combines theoretical advancements, practical applications, and educational outreach to improve mobility system design.
\end{abstract}

%%%%%%%%%%%%%%%%%%%%%%%%%%%%%%%%%%%%%%%%%%%%%%%%%%%%%%%%%%%%%%%%%%%%%%%%%%%%%%%%

\section{Introduction}
\label{sec:introduction}
In past decades, the continued rise in urbanization has had a deep impact on cities worldwide.
Currently, approximately 55\% of the world's population lives in urban areas, a figure projected to reach 70\% by 2050~\cite{un2020}.
Such population shift has increased travel demaned in metropolitan environments, exacerbating existing externalities such as congestion and pollution~\cite{czepkiewicz2018urbanites}.
As a result, cities must take complex decisions, expanding and reshaping their transportation infrastructure to accommodate new evolving travel needs, all while ensuring accessibility, equity, fairness, sustainability, and performance~\cite{ranchordas2020smart}.
In this context, the emergence of \glspl{abk:msp} such as ride-hailing operators, micromobility platforms, and, soon, \gls{abk:amod} systems~\cite{zardinilanzettiAR2021}, provides opportunities and adds complexity.
Notably, such services have gained a substantial role in urben transportation.
While they provide additional, point-to-point mobility options, they are profit-driven, and rely on public infrastructure, raising concerns about their long-term impact, regulations, and social equity~\cite{berger2018drivers,rogers2015social}.
This dynamic highlights the crucial need for effective regulatory strategies to balance public and private interests in urban mobility.
Furthermore, sustainability is a pressing priority.
Critically, cities account for approximately 78\% of the global energy consumption and over 60\% of greenhouse gas emissions~\cite{un2020bis}. Transportation alone is responsible for around 30\% of these emissions in the USA, highlighting the urgency of implementing policies that mitigate environmental impact. 
Cities and governments have recognized this challenge, and have set ambitious targets, such as New York City's plan to increase sustainable transportation modes from 68\% to 80\%, and the European Union's commitment to a 90\% emissions reduction by 2050~\cite{euplan2020,OneNYC}.

\noindent Such interconnected challenges illustrate the complexity of designing and regularing urban mobility systems, and emphasize the need for structured decision-making tools.
In this paper, we present a user-friendly \gls{abk:GUI}\footnote{Available at \href{https://bit.ly/mobility-game}{bit.ly/mobility-game}.} that leverages a simple game-theoretic framework to model multi-modal mobility systems, extending our previous efforts tackling this problem~\cite{zardinilanzetti2021,zardinilanzetti2023,zardini2023camod}.
The \gls{abk:GUI} allows users to make decisions from the perspective of various mobility system stakeholders and assess the impact of such decisions at the system level.
Furthermore, we present numerical examples for the city of Boston to illustrate its features, and provide guidelines for readers interested in replicating the results.

\paragraph{Related Literature}
Our work lies at the interface of game-theoretic modeling of transportation systems and policy making for future mobility.
Game theory has been used to formulate and solve various mobility-related problems, ranging from defining pricing strategies for \glspl{abk:msp}~\cite{mingbao2010pricing, gong2014analysis, chen2016management, kuiteing2017network, bimpikis2019spatial, yang2019subsidy, seo24tcns}, to the analysis of interactions between \glspl{abk:msp} and users~\cite{lei2018evolutionary, dandl2019autonomous, turan2021competition}, and the interactions between authorities and \glspl{abk:msp}~\cite{hernandez2018game, di2019unified, mo2021dynamic, balac2019modeling, lanzetti2023interplay}.
Further efforts have focused on competition between \glspl{abk:msp} and public transit, highlighting the potential disruptive effects of emerging mobility systems~\cite{krichene2017stackelberg, lazar2020optimal}.
Another key research direction studies the role of game theory in policy making, specifically in managing congestion and designing regulatory interventions, such as tolling strategies~\cite{bianco2016game}, incentive mechanisms~\cite{swamy2012effectiveness, paccagnan2021optimal, lazar2020optimal}, and Stackelberg games for traffic network management~\cite{krichene2017stackelberg}.

In addition to game-theoretic approaches, policy research on emerging mobility systems has focused on mitigating externalities and achieving socially efficient solutions.
In this context, studies have proposed mechanisms such as congestion pricing, labor subsidies, and land-use regulations to reduce transportation-related emissions~\cite{fullerton2002can, iwata2016can, zhang2016optimal, maljkovic2023hierarchical}.
Furthermore, economic analyses of ride-hailing and carpooling models have been proposed~\cite{zoepf2018economics,ostrovsky2019carpooling}.
However, while these studies provide valuable insights into specific aspects of mobility policy, they often focus on isolated stakeholders, or ignore the interactive dynamics between different players in the mobility ecosystem.
Our work aims to bridge this gap by developing a structured, game-theoretic approach accounting for multi-level decision-making in transportation systems.
\paragraph{Statement of Contribution}
In this paper, we build on~\cite{belgioioso2021semi, zardini2021game, zardini2022} to develop a game-theoretic formulation of interactions in multi-modal mobility systems.
Specifically, we present a framework to formulate interactions problems and solve them to find equilibria.
Furthermore, we present a \gls{abk:GUI} to be used as an educational tool to visualize the complexity of the problem, as well as decision-making tool for stakeholders of the mobility ecosystem. 
Finally, we showcase the framework via a case study of Boston/Cambridge, Massachusetts, USA. 
We conduct a sensitivity analysis with real-world scenarios to highlight related complexities.
\paragraph{Organization of the Manuscript}
The remainder of this paper is organized as follows. 
We specify the problem setting and models for interactions in mobility systems in \cref{sec:model}.
In \cref{sec:results} we detail case studies and present numerical results.
Finally, we draw conclusions and present future research interests in \cref{sec:conclusion}.
\section{Modeling Interactions in Multi-Modal Mobility Systems using Population Games}
\label{sec:model}
% Note for myself later (for the process): method for writing the section, read through the code and made sure all the variables in the code are accounted for in the model
In this section, we present a model characterizing the interactions between stakeholders of a multi-modal mobility ecosystem.
Specifically, our model builds on~\cite{belgioioso2021semi}, which focuses on solving generalized Nash Equilibrium problems in monotone aggregate games.

\subsection{Modeling of the Populations and Zones}
We model a city by means of~$N$ zones, indexed by~$i \in \{1, \dots, N\}$. 
Each citizen in zone~$i$ seeks to travel to a different zone~$j$ where~$j \in  \{1, \dots, N\} \setminus \{i\}$ (i.e., we do not allow any citizen to stay in their original zone).
To travel between zones, citizens can choose between~$M$ modes of transportation.
Rather than modeling the behavior of each individual separately, we group travelers into~$K$ distinct populations (e.g., students, employees, and leisure travelers).
For population~$k$ in zone $i$, we denote~$d_{ijk}$ as the travel demand for population~$k$, representing the number of individuals from population~$k$ who wish to travel from zone~$i$ to zone~$j$.
By definition,~$d_{ijk} \in \mathbb{R}_{\geq 0}$ for all $i,j \in \{1, \dots, N\}$ and $k \in \{1, \dots, K\}$.
\subsection{Modeling of the MSPs}
In this model, we focus on the capacity of \glspl{abk:msp} as they depart from the origin zone.
Each provider~$m$ has a maximum capacity~$C_i^m$ for trips originating in zone~$i$.
Importantly, note that~$C_i^m$  represents the total number of available ``seats'', and not the number of actual vehicles.
This distinction is important for modes of transport where a single vehicle can carry multiple passengers per trip (e.g., buses).
The value of~$C_i^m$ may depend on various factors, such as the municipal budget for purchasing vehicles, or the cost of licenses for \gls{abk:amod} systems.
In the developed \gls{abk:GUI}, however,~$C_i^m$ is determined by the number of vehicles allocated to each zone by the player.
In this model, we do not distinguish capacity for various destinations that originate from the same origin.
Instead, we consider the maximum number of passengers that mode~$m$ can transport from the origin zone as a whole.
Finally, we associate the index~$m=0$ to walking, and set~$C_i^0 = +\infty $ for all~$i \in N$, ensuring that walking is always a feasible mode of transportation (the slowest one). 
\subsection{Cost of Traveling}
The cost~$c_{ijk}^m$ for an individual from population~$k$ traveling from zone~$i$ to zone~$j$ via mode~$m$ is determined by combining the monetary fare for the selected mode with the travel time~$t_{ij}^m$.
The travel time is converted into monetary terms leveraging the average value of time for population~$k$, reflecting the opportunity cost of travel.
For each origin-destination (OD) pair~$i,j$ and mode~$m$, the travel time~$t_{ij}^m \in \mathbb{R}_{\geq 0}$ represents the time required to travel from zone~$i$ to zone~$j$.
If~$i=j$, then~$t_{ij}^m=0$ for all modes.
Notably,~$t_{ij}^m$ and~$t_{ji}^m$ are not necessarily equal; for instance, during morning rush hour, commuting into a city may take longer than leaving it.

\noindent Let~$p_{ij}^m$ denote the monetary fare for traveling from zone~$i$ to zone~$j$ using mode~$m$, and let~$w_k$ represent the average value of time for population~$k$ (e.g., calculated based on the average wage of a citizen of population~$k$ inn a given city).
The cost for an individual is then given by
\begin{equation}\label{cost fxn}
    c_{ijk}^m = p_{ij}^m + w_k \cdot t_{ij}^m.
\end{equation}
Notably, the costs experienced by individuals are independent of the number of other travelers (e.g., the cost of taking a bus is the same whether it is shared with one person or fifty).
For the special case of walking ($m=0$), we define the fare as~$p_{ij}^0=0$, ensuring that walking is alwas a feasible option.
However, note that the total cost of walking is not necessarily zero, as it is more time-intensive than other modes of travel.

\subsection{Equilibrium}
We represent the decisions of each population using the variables~$\{x_{ijk}^m\}$, where~$x_{ijk}^m \in [0,1]$ denotes the proportion of population~$k$ along route~$ij$ traveling via mode~$m$.
The total number of citizens in population~$k$ is denoted by~$P_k$, and~$P_{i,j,k}$ represents the number of citizens from population~$k$ traveling along route~$ij$. 
For a configuration to be \emph{feasible}, it must meet the transportation demand while adhering to system's capacity constraints.
\begin{definition}[Feasible configuration] \label{def: feasibility}
A configuration~$\{x_{ijk}^{m}\}$ is \emph{feasible} if it satisfies the following conditions:
\begin{enumerate}
\item $x_{ijk}^m\geq 0$ for all~$i,j,k,m$;
\item No citizen of population~$k$ may remain in their original zone, i.e.,
\begin{equation*}
    \sum_{m}x_{iik}^m=0, \text{ for all }i,k;
\end{equation*}
\item All citizens of population~$k$ traveling along route~$ij$ must reach their destination, i.e.,
\begin{equation*}
    \sum_{m}x_{ijk}^m=1, \text{ for all }i,j,k;
\end{equation*}
\item The total demand across all origin-destination pairs and modes must equal the population size, i.e.,
\begin{equation*}
    \sum_{i,j,m} d_{ijk}x_{ijk}^m=P_k \text{ for all }k;
\end{equation*}
\item The system's capacity limits must not be exceeded, i.e., 
\begin{equation*}
    \sum_{j,k} d_{ijk} x_{ijk}^m \leq C_{i}^m \text{ for all }i,m;
\end{equation*}
\end{enumerate}
\end{definition}

From this definition, we have established the minimum requirement for an \emph{equilibrium}. Now, we establish the requirement for an \emph{optimal} solution. A \emph{Nash equilibrium} is attained when no agent can lower their travel cost by taking another mode of transportation. Formally: 

\begin{definition}[Nash Equilibrium]
Let~$\{x_{ijk}^{m}\}$ be a feasible configuration.
We say that~$\{x_{ijk}^m\}$ is a \emph{Nash equilibrium} of the game if for all~$i,j,k,m$ with~$x_{ijk}^{m}>0$ every other mode~$m'\in\{0,\ldots,M\}$ either (i) leads to higher cost, i.e., 
\begin{equation*}
c_{ijk}^m\left(\sum_{j,k}  P_{i,j, k} x_{ijk}^m\right)
    \leq 
c_{ijk}^{m'}\left(\sum_{j,k}  P_{i,j,k} x_{ijk}^{m'}\right),
\end{equation*}
or (ii) it is saturated, i.e., 
\begin{equation*}
    \sum_{j,k}  P_{i,j,k} x_{ijk}^m = C_{i}^m.
\end{equation*}
\end{definition}

In other words, a Nash equilibrium occurs when no subset of individuals can reduce their individual travel cost by switching modes, assuming that all others maintain their choices.
Furthermore, the system-wide travel cost at equilibrium is locally optimal, meaning that no alternative configuration reduces the cost without violating the feasibility constraints.

\subsection{Analysis of the Game}
In this section, we show that the game can be analyzed as a convex optimization problem, guaranteeing \emph{existence} of equilibria. Our result takes the following form:

\begin{theorem}[Equilibria of the game]\label{thm:equilibria}
Let~$\{x_{ijk}^{m}\}$ be a feasible configuration resulting from the convex optimization problem
\begin{subequations}\label{eq:optimization}
\begin{align}
    \min_{x_{ijk}^m}\: &
    \sum_{i,j,k,m} c_{ijk}^m  P_{i,j,k} x_{ijk}^m 
    \\
    \mathrm{s.t. }\: &
    \sum_{j,k} d_{ijk} x_{ijk}^m \leq C_{i}^m \text{ for all }i,m
    \\
    & \sum_{m}x_{ijk}^m=1, \text{ for all }i,j,k, 
    \\
    & \sum_{i,j,m} d_{ijk}x_{ijk}^m=P_k \text{ for all }k
    \\
    &x_{ijk}^m\geq 0 \text{ for all }i,j,k,m.
\end{align}
\end{subequations}
Then,~$\{x_{ij}^{k,m}\}$ is an equilibrium. 
In particular, an equilibrium \emph{always} exists.
\end{theorem}

\begin{remark}[Convexity and Linearity]
    The optimization problem~\eqref{eq:optimization} is indeed convex. 
    All constraints are clearly linear in the decision variables. 
    To show convexity of the objective function, observe that the cost of traveling for each citizen is independent. We aim to minimize the cost of traveling for all agents, but because costs are independent, minimizing the total cost is the same as minimizing the individual cost functions. This makes the objective function also linear in the decision variables. Since the optimization problem is linear, it is also convex. 
    In particular, the optimization problem has~$N^2MK$ decision variables and~$\mathcal{O}(N(N-1)MK)$ constraints (since~$x_{iik}^m=0$ for all~$i\in\{1,\ldots,N\}$).
\end{remark}

\cref{thm:equilibria} states that we can solve for equilibria of the game using an off-the-shelf convex optimizer. In the case of the \gls{abk:GUI}, since our problem is linear, we used the \texttt{linprog} function from SciPy's optimization module.

\subsection{Discussion}
It is worth noting a few key points. First, we abstract away from the topology of the road and public transit network and disregard personal vehicles. By doing so, we can assume that all travelers have instant access to their mode of choice and there is no congestion from other vehicles (public or private). Second, the model cannot account for repeated operation of the system within the given time frame. The \gls{abk:GUI} asks the user to set parameters for a one hour timeframe, and it models all trips within that hour simultaneously. 
\section{Case studies}
\label{sec:results}
For our case study, we consider Boston and Cambridge, Massachusetts, USA, as a unified system. Together, these cities have a combined population of approximately 750,000 and cover an area of 55 square miles~\cite{area_bos,area_cam}. 
In this study, we analyze the travel behavior of around 30,000 citizens per hour, representing approximately~$4\%$ of the population in motion at any given time. 
To structure the study, the region is divided into eight zones, each centered around key city landmarks: Massachusetts Institute of Technology, Harvard University, Massachusetts General Hospital, Logan Airport, Boston City Hall, Boston Common, the Prudential Center, and Fenway Park. 
These landmarks were selected to capture a mix of essential services (airport, hospital, universities, government buildings) and leisure destinations (parks, shopping centers, and entertainment venues).
It is important to note that this model does not capture all travel movements in the city.
Specifically, residential and office areas are not explicitly included in the zone definitions.
As commuting between home and work accounts for a significant portion of daily travel, and such trips typically follow consistent patterns less influenced by cost, we exclude these travelers from our analysis. 
Instead, our focus is on more dynamic travel behavior, where mode choice is more sensitive to pricing and availability.
\noindent Without loss of generality, we categorize travelers into three distinct populations, based on trip intent: employees, students, and leisure travelers. 
Note that a traveler's classification depends on the purpose of their trip (e.g., a working professional commuting to the office is classified as an employee, but the same individual visiting a park on the weekend is considered a leisure traveler).
The distribution of travelers originating from each zone is determined by the nature of its defining landmark.
Hospitals, for instance, generate a high number of employee trips, whereas universities primarily contribute to student travel.
In particular, the data on trip patterns is sourced from local survey.
We consider four modes of transportation: bus, \gls{abk:amod}, shared bikes, and walking. 

\begin{figure}[h]
    \centering
    \includegraphics[width=0.9\linewidth]{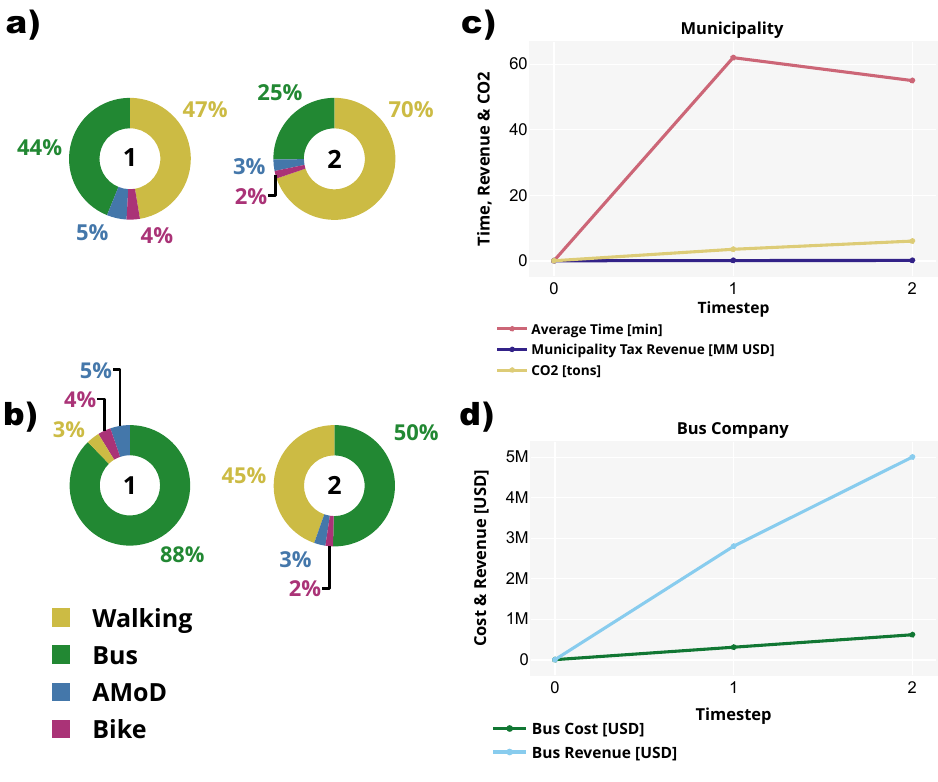}
    \caption{a) shows the mode-share proportion of citizens leaving zones~$1$ and~$2$. b) After doubling the number of buses in each zone from~$15$ to~$30$, the proportion of citizens traveling via bus also doubles. A mode-shift from walking to bus can be observed at the second iteration. c) Line graph showing the average travel time, municipality revenue from bus and bike tax, and CO$_2$ emissions produced during each timestep. When doubling the number of buses, the average travel time decreases and CO$_2$ emissions increase. d) Line graph showing the revenue and costs incurred by the bus company. When doubling the number of buses, costs marginally increase, but revenue significantly increases.}
    \label{fig:double_num_buses}
\end{figure}

\noindent Using the \gls{abk:GUI}, users can configure the transportation system from the perspective of the municipality and three private mobility providers (the bus company, the \gls{abk:amod} operator, and the bike-sharing service). 
When acting as the municipality, users set tax rates on bikes and cars as a percentage of company revenue. 
As private mobility \glspl{abk:msp}, users control fleet allocation across zones and determine pricing strategies. 
Bus fares are per trip, whereas \gls{abk:amod} and bike-sharing fares are distance-based.
The system reaches an equilibrium by minimizing travel costs while respecting capacity constraints. 
To assess the quality of the equilibrium, we report several metrics of interest, including average travel time for a citizen across all routes, modes, and populations, environmental impact in form of CO$_2$ emissions for the various modes, and costs and revenues for \glspl{abk:msp}.
The \gls{abk:GUI} enables users to iteratively modify the city's configuration and compare the resulting equilibria. 
In the following, we provide a detailed overview of the \gls{abk:GUI}'s functionality and its role in facilitating system analysis and showcasing the complexity of the problem.

\subsection{GUI Working Principle}
We conduct case studies and analyze the results based on the visualization provided by the \gls{abk:GUI}. 
First, we present a nominal case to introduce the core functionalities of the \gls{abk:GUI}. 
Then, we examine how changes in the number of buses and the pricing of \gls{abk:amod} services affect equilibrium outcomes by adjusting the corresponding parameters and comparing the visualized results.

\begin{figure}[h]
    \centering
    \includegraphics[width=0.7\linewidth]{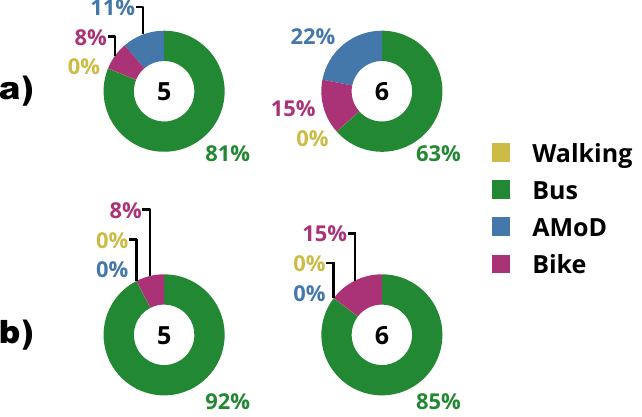}
    \caption{a) shows the mode-share proportion of citizens leaving zones~$5$ and~$6$. b) After doubling the price of \gls{abk:amod}, the proportion of citizens traveling via \gls{abk:amod} decreases to~$0\%$ due to citizens choosing a cheaper mode, which is primarily bus.}
    \label{fig:double_amod_price}
\end{figure}

\begin{figure*}[t]
    \includegraphics[width=\linewidth]{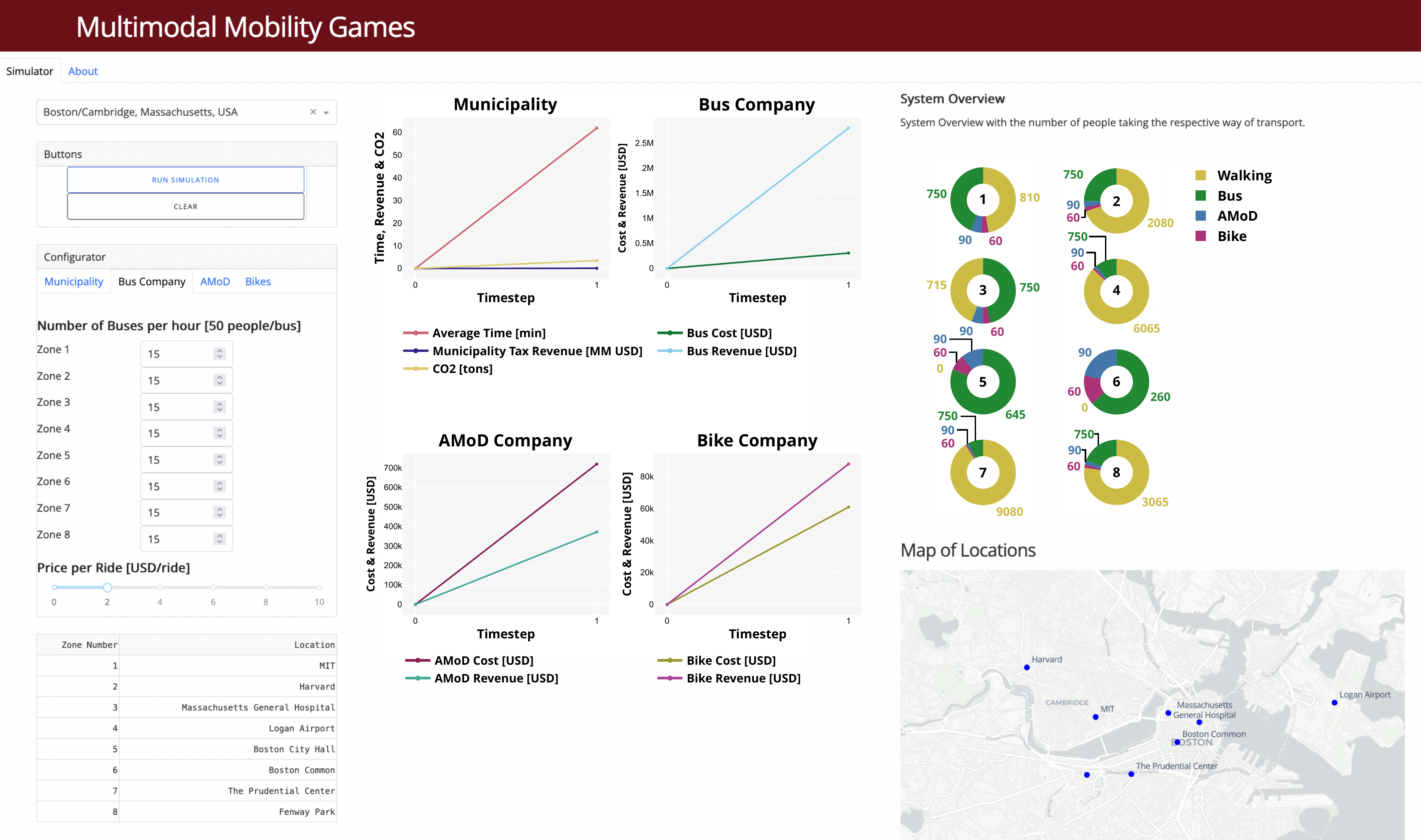}
    \caption{Overview of the \gls{abk:GUI}. In the leftmost column is the drop down menu for city selection, the buttons to run and re-run the simulation, an interaction menu to set parameters, and a table enumerating the zones in the city. In the center column, the user is presented with graphs showing performance metrics across simulation iterations. In the rightmost column, a ``System Overview" visualizing the proportion of all citizens \textit{leaving} zone~$i$ traveling by mode~$m$. Finally, a map allows the user to locate the studied zones geographically.}
    \label{fig:gui_overview}
\end{figure*}

\paragraph*{Nominal Case}
To establish a baseline scenario, we place~$15$ buses,~$90$ AMoD vehicles, and~$60$ bikes in each of Boston's~$8$ zones. 
These values are derived from the total daily availability of buses~\cite{num_buses}, taxis~\cite{num_taxis} and Ubers~\cite{num_ubers}, and Blue Bikes~\cite{num_bikes} in Boston. 
% source buses: http://roster.transithistory.org/
% source bikes: https://www.boston.gov/departments/transportation/bluebikes
% source taxi: https://www.welcomepickups.com/boston/taxi/
% source uber: https://www.americaninno.com/boston/uber-driver-data-boston-has-10000-uber-drivers/
We impose a 20\% revenue tax on cars and bikes, while bus fares remain fixed at 2 USD per ride~\cite{busfares}.
\gls{abk:amod} and bike-sharing fares are distance-based, with \gls{abk:amod} services starting at 1 USD per mile, and bike-sharing at 0.20 USD per mile.
% bus fare source: https://www.mbta.com/fares/bus-fares
% amod numbers from finding uber price and dividing by distance
%
In line with existing studies~\cite{lanzetti2023interplay}, we define the value of time of employees, students, and leisure travelers as~$35$ USD/h,~$15$ USD/h, and~$7$ USD/h, respectively. 
Using these parameters, the \gls{abk:GUI} generates the output reported in \cref{fig:gui_overview}.
Based on the nominal case shown in \cref{fig:gui_overview}, we make two adjustments to observe how the system reacts. 
First, we double the number of buses per zone from~$15$ to~$30$. 
Second, we double the price of \gls{abk:amod} from~$1$ USD per mile to~$2$ USD per mile. In each case, we hold all the other variables constant from the nominal case.

\paragraph*{Doubling the number of buses}
In this iteration, we double the number of buses per zone from~$15$ to~$30$, holding all other parameters constant. 
Focusing on zones~$1$ and~$2$, the original bus capacity was~$750$ passengers per zone, which accounted for~$44\%$ mode-share (see~\cref{fig:double_num_buses}a).
After increasing the number of buses, the capacity doubles to 1,500 passengers per zone, and bus mode-share doubles (see~\cref{fig:double_num_buses}b).
This shift results in several notable effects.
First, since walking is the slowest mode, replacing it with bus transit significantly shortens travel duration for affected travelers.
Second, a higher number of buses leads to greater fuel consumption and emissions (see~\cref{fig:double_num_buses}c).
Finally, more passengers generate additional fare-based income for the bus service (see~\cref{fig:double_num_buses}d).

% \begin{figure}[t!]
% \centerline{\includesvg[width=0.75\columnwidth]{CCTA25/img/pixel_perfect_fig2.svg}}
% \caption{Example of using SVG on Overleaf}
% \label{fig: example}
% \end{figure}

% \includesvg[width=0.5\textwidth]{CCTA25/img/pixel_perfect_fig2}

% \begin{figure}[h]
%     \centering
%     {\includesvg[width=7cm]{CCTA25/img/pixel_perfect_fig2.svg}}
%     \caption{test fig2}
%     \label{fig:iter1_zones12}
% \end{figure}

% \begin{figure}[h]
%     \centering
%     \includegraphics[width=7cm]{CCTA25/img/iter1_zones12.png}
%     \caption{first iteration: this should be one fig with the second iteration pie charts?}
%     \label{fig:iter1_zones12}
% \end{figure}

% \begin{figure}[h]
%     \centering
%     \includegraphics[width=7cm]{CCTA25/img/iter2_zones12.png}
%     \caption{second iteration: this should be one fig with the first iteration pie charts?}
%     \label{fig:iter2_zones12}
% \end{figure}

% \begin{figure}[h]
%     \centering
%     \includegraphics[width=7cm]{CCTA25/img/iter2_graphs_mun_bus.png}
%     \caption{this should be one fig with the pie charts? and i need the legend}
%     \label{fig:fig2_graph}
% \end{figure}

\paragraph*{Doubling the price of \gls{abk:amod}}
In this iteration, we double the price of \gls{abk:amod} from~$1$ USD per mile to~$2$ USD per mile. 
In the first iteration, \gls{abk:amod} was fully saturated due to its significantly higher speed compared to other modes (see~\cref{fig:double_amod_price}a).
When fares across modes are similar, travel time becomes the dominant factor in decision-making.
However, with the price increase to 2 USD per mile, the cost of \gls{abk:amod} rises, making it less attractive to travelers.
As a result, many citizens switch to alternative modes, primarily buses.
This shift is particularly evident in zones 5 and 6, as shown in~\cref{fig:double_amod_price}b, where a substantial portion of former \gls{abk:amod} users opt for bus transit instead.
% \begin{figure}[h]
%     \centering
%     \includegraphics[width=7cm]{CCTA25/img/iter1_zones56.png}
%     \caption{first iteration: this should be one fig with the second iteration pie charts?}
%     \label{fig:iter1_zones12}
% \end{figure}

% \begin{figure}[h]
%     \centering
%     \includegraphics[width=7cm]{CCTA25/img/iter2_zones56.png}
%     \caption{second iteration: this should be one fig with the first iteration pie charts?}
%     \label{fig:iter2_zones12}
% \end{figure}

\subsection{Remarks}
The demand patterns used in these case studies are approximate or rounded estimates.
Going forward, incorporating more granular and accurate data would allow for a more precise representation of the city's mobility dynamics.
In this analysis, we modified one variable a a time to isolate its impact and demonstrate how the model responds to specific parameter changes.
However, simultaneously altering multiple variables can introduce unexpected network-wide effects, underscoring the complexity and interdependencies inherent in urban mobility systems.

\subsection{\gls{abk:GUI} capabilities}
In this paper, we have shown two iterations for one region with eight zones.
The \gls{abk:GUI} currently supports 3 realities: Lugano, Switzerland ($8$ zones); 
Boston/Cambridge, MA, USA ($8$ zones); and Kyiv, Ukraine ($12$ zones). 
Importantly the \gls{abk:GUI} is designed with a strong emphasis on user engagement, allowing users to contribute and model their own cities.
The user must provide coordinates of the zones, the number of citizens per population at each zone, demand patterns along each route for each population, and information about value of time for each population.

\section{Conclusion}\label{sec:conclusion}
In this paper we have studied the impact of transportation policies and mobility service design leveraging a simple, game-theoretic framework integrated into an interactive \gls{abk:GUI}.
Specifically, we modeled urban mobility dynamics in Bostom/Cambridge.
Through a series of case studies, we examined how changes in fleet allocation, pricing, and taxation influence equilibrium travel behavior, modal choices, and efficiency.
By iterating on these properties, we highlighted the complex interdependencies within mobility systems, reinforcing the need for data-driven decision-making in urban transportation planning. 
Importantly, the \gls{abk:GUI} serves as both as a research tool and an educational platform, allowing users to configure cities, simulate transportation scenarios, and analyze network-wide effects.
Importantly, it provides an engaging way for students and policymakers to explore the challenges of urban mobility, gaining hands-on experience in balancing efficiency, sustainability, and financial factors.

Future work will focus on improving the model capabilities, incorporating more granular, real-world data, and refining modeling techniques. 
\section{Acknowledgments}
We want to thank Christian Hartnik, Nicolas Lanzetti, Saverio Bolognani, and Giuseppe Belgioioso, who contributed to an earlier version of the GUI and project.

%\section*{Acknowledgments}
%\ldots

%%%%%%%%%%%%%%%%%%%%%%%%%%%%%%%%%%%%%%%%%%%%%%%%%%%%%%%%%%%%%%%%%%%%%%%%%%%%%%%%
\bibliography{paper}

%\clearpage
%\appendix
%\input{chapters/appendix}
%\bibliography{paper.bbl}

\end{document}